\documentclass[conference]{IEEEtran}
\IEEEoverridecommandlockouts
\usepackage{cite}
\usepackage{amsmath,amssymb,amsfonts}
\usepackage{graphicx}
\usepackage{textcomp}
\usepackage{multirow}
\usepackage{algpseudocode}
\usepackage{microtype}
\usepackage{marvosym}
\usepackage{bm}
\usepackage{url}

\usepackage{xcolor}
\def\BibTeX{{\rm B\kern-.05em{\sc i\kern-.025em b}\kern-.08em
    T\kern-.1667em\lower.7ex\hbox{E}\kern-.125emX}}
\begin{document}

\title{Gender Bias in Depression\linebreak Detection Using Audio Features}

\author{\IEEEauthorblockN{Andrew Bailey, Mark D. Plumbley}
\IEEEauthorblockA{\textit{Centre for Vision, Speech and Signal Processing} \\
\textit{University of Surrey}\\
Guildford, U.K. \\
\{andrew.bailey, m.plumbley\}@surrey.ac.uk}
}

\maketitle

\begin{abstract}
Depression is a large-scale mental health problem and a challenging area for machine learning researchers in detection of depression. Datasets such as Distress Analysis Interview Corpus - Wizard of Oz (DAIC-WOZ) have been created to aid research in this area. However, on top of the challenges inherent in accurately detecting depression, biases in datasets may result in skewed classification performance. In this paper we examine gender bias in the DAIC-WOZ dataset. We show that gender biases in DAIC-WOZ can lead to an overreporting of performance. By different concepts from Fair Machine Learning, such as data re-distribution, and using raw audio features, we can mitigate against the harmful effects of bias.
\end{abstract}


\section{Introduction}
Depression is a mental health disorder that causes severe symptoms for individuals. Typical emotional symptoms include lasting feelings of unhappiness, hopelessness, and a lack of interest or enjoyment. The speech of individuals with depression has been characterised as lifeless or flat, and delivered in a monotone fashion \cite{scherer_self-reported_2016}. Depression is a common and costly mental health disorder. McManus et al. \cite{mcmanus_mental_2016} reported in 2014 that $3.8\%$ of adults in England suffered from depression and, according to the OECD \cite{oecd_health_2018}, the cost of mental health in the UK was around $4\%$ of GDP (\EUR{106}bn) in $2015$. 

Machine learning techniques have successfully been applied to many health-related areas \cite{sendak_human_black_box_2020, peimankar_ensemble_2019} and therefore have the potential to improve depression detection. The Distress Analysis Interview Corpus - Wizard of Oz (DAIC-WOZ) dataset \cite{gratch_distress_2014} was designed to facilitate research into depression detection and was released as part of the 2016 Audio-Visual Emotion Challenge (AVEC) \cite{valstar_avec_2016}. DAIC-WOZ contains interviews from 189 participants: 57 diagnosed with post-traumatic stress disorder or depression, and 132 who are not.

The ground truth labels in DAIC-WOZ were obtained through a depression diagnosis questionnaire, the Patient Health Questionnaire 8 (PHQ-8) \cite{kroenke_phq-8_2009}, which consists of $8$ questions. Every participant in the DAIC-WOZ answered the $8$ questions by assigning an integer score in the range $0-3$. Once all the questions were completed, the scores were added together to give a depression rating, $r$ out of $24$, where $r<10$ is classified as not depressed and $r\geq 10$ is classified as depressed.

Traditional machine learning algorithms such as Support Vector Machines (SVM) were used by the AVEC 2016 \cite{valstar_avec_2016} organisers as an audio baseline classification system. Since then, deep learning methods have also been applied to DAIC-WOZ including Convolutional Neural Networks (CNN) \cite{dubagunta_learning_2019, ma_depaudionet:_2016, mdhaffar_dl4ded_2019}, Fully Connected Deep Neural Networks \cite{yang_multimodal_2017}, and Recurrent Neural Networks (RNN) \cite{ma_depaudionet:_2016, mdhaffar_dl4ded_2019}, along with hand-crafted features, including the spectrogram \cite{mdhaffar_dl4ded_2019} and the mel-spectrogram \cite{ma_depaudionet:_2016}.

However, when machine learning is used to tackle health-related problems, such as depression detection, considerations must be taken to ensure that these models do not incorporate bias \cite{hebert-johnson_multicalibration_2018}. Bias can result in subjects with protected characteristics, such as race or gender, being unfairly penalised. This can occur due to many factors such as poor data collection standards \cite{yang_towards_2020} or the data processing in machine learning models \cite{yang_towards_2020}. 

Fair Machine Learning (Fair ML) \cite{kearns_empirical_2019} is an area of research that explores the idea of fair, unbiased classification, in an attempt to work towards a fairer society through the use of machine learning. Fair ML techniques have been used in healthcare domains such as the paralinguistics community to detect dementia \cite{paralinguistics}. 

We show in this paper that the DAIC-WOZ dataset contains gender bias and that this bias can negatively affect the resulting accuracy of machine learning models using audio features. We also find that deep learning models based on raw audio are more robust to gender bias than ones based on other common hand-crafted features, such as mel-spectrogram. 

Our paper is organised as follows. Section \ref{sec:daic-woz, depression detection, and fairness} explores the background on the DAIC-WOZ dataset and Fair ML, Section \ref{sec:depaudionet reproduction} discusses the baseline model architecture, Section \ref{sec:proposed method} explores our proposed method, Section \ref{sec:experimentation with gender balance} explores the bias problem and how to mitigate against it and Section \ref{sec:conclusion} concludes the paper.  

\section{DAIC-WOZ, Depression Detection, and Fairness}
\label{sec:daic-woz, depression detection, and fairness}
\subsection{DAIC-WOZ Dataset}
\label{subsec:daic-woz dataset}

\begin{figure}
  \centering
  \includegraphics[width=2.5in]{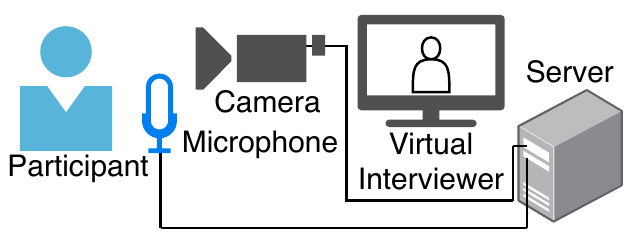}
  \caption{A typical interview scenario where a participant converses with virtual interviewer on a TV screen. The camera records the facial movements and the microphone records the audio of the session.}
  \label{fig:daic-woz-interview}
\end{figure}

To gather the data for the DAIC-WOZ dataset, interviews were conducted between each participant and a virtual interviewer (“Ellie”) controlled by a researcher in another room \cite{SimSensei}. The audio and facial features of the participants were recorded (Figure \ref{fig:daic-woz-interview}). The interviews range from 7 minutes to 35 minutes. 

For Machine Learning purposes, the DAIC-WOZ dataset is divided into a training set of 107 files (76 non-depressed (ND) interviews and 31 depressed (D) interviews), a validation set of 35 files (23 ND interviews and 12 D interviews), and a test set of 47 files. The distribution of data for the test set is hidden due to the AVEC competition \cite{valstar_avec_2016}. 

\begin{table}[t]
  \caption{Subdivisions of training files (gender and class) of DAIC-WOZ.}
  \centering
  \label{tab:training_quadrants_train}
  \begin{tabular}{| c | c | c | c |} 
  \hline
   & \textbf{ND} & \textbf{D} & \textbf{Total (ND+D)}\\ 
   \hline
   Female (F) & $27\ (25\%)$ & $17\ (16\%)$ & $44\ (41\%)$ \\
   \hline
   Male (M) & $49\ (46\%)$ & $14\ (13\%)$ & $63\ (59\%)$\\ 
   \hline
   Total (F+M) & $76\ (71\%)$ & $31\ (29\%)$ & $107\ (100\%)$\\
   \hline
  \end{tabular}
\end{table}

\begin{table}[t]
  \caption{Subdivisions of validation files (gender and class) of DAIC-WOZ. Test set details are not released for competition purposes.}
  \centering
  \label{tab:training_quadrants_validation}
  \begin{tabular}{| c | c | c | c|} 
  \hline
   & \textbf{ND} & \textbf{D} & \textbf{Total (ND+D)}\\ 
   \hline
   Female (F) & $12\ (34\%)$ & $7\ (20\%)$ & $19\ (54\%)$ \\
   \hline
   Male (M) & $11\ (31\%)$ & $5\ (14\%)$ & $16\ (46\%)$\\ 
   \hline
   Total (F+M) & $23\ (65\%)$ & $12\ (34\%)$ & $35\ (100\%)$\\
   \hline
  \end{tabular}
\end{table}

To maintain the privacy of the participants, the raw visual data is not released. Instead, visual features based on the OpenFace framework \cite{baltrusaitis_openface:_2016} and the FACET toolbox \cite{littlewort_computer_2011} were extracted. Raw audio files, sampled at 16 kHz, are provided.

\subsection{Gender Bias in DAIC-WOZ}
\label{subsec:the daic-woz data imbalance}
As previous researchers have noted, we can see from Table \ref{tab:training_quadrants_train}, that there is a high-level class imbalance of roughly $2$:$5$ in terms of D:ND interviews in the training set. 

In addition, we also see a gender bias in Table \ref{tab:training_quadrants_train} that has not been explicitly acknowledged by previous researchers. While the ratio of D:ND females is roughly $5$:$8$, the ratio of D:ND males in the dataset is $2$:$7$. Thus the frequency of individuals with depression in the dataset differs between genders, such that \(p(D\ |\ g=\rm{f}\)\() > p(D\ |\ g=\rm{m}\)\()\) where $g$ is gender, $\rm{f}$ is female and $\rm{m}$ is male. We also see a similar class imbalance as well as gender bias in the validation set in Table \ref{tab:training_quadrants_validation}.

\subsection{Training on Separate Gender Subsets}
\label{subsec:training on seperate gender subsets}
Based on research indicating that depressive symptoms can differ depending on gender \cite{honig_automatic_2014}, some authors split DAIC-WOZ into two gendered subsets, one containing the female interviews and one containing the male interviews \cite{cummins_enhancing_2017, vlasenko_implementing_2017, yang_multimodal_2017}. These authors \cite{cummins_enhancing_2017, vlasenko_implementing_2017, yang_multimodal_2017} did not report awareness of the gender bias.

While these authors reported improved performance detecting depression on DAIC-WOZ compared to not splitting the dataset, deep learning benefits from large amounts of data, therefore training two separate models on subsets of the larger training data may may lead to reduced performance. 

In addition, gender differences in speech signals are well known \cite{ishikawa_toward_2017}. Huang et al. \cite{huang_natural_2020} proposed to reduce gender dependency in speech signals by applying z-normalisation to every audio signal based on gender according to, \(\Tilde{x} = (x - \mu{})/\sigma\), where $x$ is the input feature, $\mu$ and $\sigma$ are the mean and standard deviation of the features in the DAIC-WOZ training set split by gender and $\Tilde{x}$ is the normalised feature. Huang et al. \cite{huang_natural_2020} did not compare results before and after gender normalisation and did not comment on the larger gender bias in the dataset.  

In this paper, we will discuss ways to deal with gender bias in the training set without creating two gendered subsets. To do this, let us explore some recent work on fair machine learning to resolve gender bias.

\subsection{Fair Machine Learning}
\label{subsec:fair machine learning}
In Fair Machine Learning (Fair ML), we are interested in making sure that protected characteristics, such as race or gender, are treated equally. One Fair ML approach, \textit{statistical parity} \cite{barocas-hardt-narayanan}, is designed to ensure that an acceptance rate, such as admission to a University programme, is the same regardless of the protected characteristics of an individual, i.e.:

\begin{equation}
\label{eq:statistical_parity}
    \mathop{\mathbb{P}}\ [R=1\ |\ A=a] = \mathop{\mathbb{P}}\ [R=1\ |\ A=b]
\end{equation}
\noindent where $R$ is the score and $A$ is a protected characteristic. 

Another principle that Fair ML follows, known as \textit{sufficiency}, says a target label $Y=1$ is independent of a protected characteristic, $A$, given some score $R=r$, i.e.:
\begin{equation}
\label{eq:protected group effect on classifier}
    \mathop{\mathbb{P}}\ [Y=1\ |\ R=r,\ A=a] = \mathop{\mathbb{P}}\ [Y=1\ |\ R=r,\ A=b]
\end{equation}
An example of \textit{sufficiency}, there should be an equal probability that two individuals actually have cancer given they received the same score regardless of their protected characteristic. 

Krawczyk \cite{krawczyk_learning_2016} mentions several approaches for ensuring an unbiased classifier, including data generation techniques such as data sub-sampling, the process of equalising the number of examples from each class in the dataset by randomly selecting a portion of examples from the majority classes. 

For more examples of fairness criteria and alternative definitions, additional work can be found at \cite{liu_implicit_2019, chouldechova_fair_2017, hardt_equality_2016}. 

In this paper, we will determine the effects of gender bias in the DAIC-WOZ dataset on a deep learing model and apply data sub-sampling techniques from Krawczyk \cite{krawczyk_learning_2016} to tackle this bias.

\section{DepAudioNet}
\label{sec:depaudionet reproduction}
Our model will be based on ``DepAudioNet'' of Ma et al. \cite{ma_depaudionet:_2016}, which was one of the best models in AVEC 2016 to utilise deep learning with audio-only processing for depression detection. 

\subsection{File Processing}
\label{subsec:File Processing}
DepAudioNet \cite{ma_depaudionet:_2016} extracts mel-spectrogram features (mel filter bank with $40$ frequency bins) using a Hanning window of length $w=1024$, and hop size $h=512$. Next, the features from each audio signal are normalised according to the z-normalisation ($\Tilde{x}=(x-\mu{})/\sigma$) where $\Tilde{x}$ is the normalised feature, $x$ is the feature from the input audio signal, $\mu$ and $\sigma$ are the mean and standard deviation calculated based on the features from each individual audio signal.

Ma et al. \cite{ma_depaudionet:_2016} randomly crop every normalised feature to the same size as the shortest audio signal. To account for the DAIC-WOZ D:ND class imbalance (ND $>$ D), they then randomly sub-sample the ND features to train with an equal number of D and ND examples. They then split the resulting feature into multiple temporal segments of size $N_{\rm{seg}}=120$ to be used as input for DepAudioNet.

\subsection{Model Architecture}
\label{subsec:model architecture}

\begin{figure}
  \centering
  \includegraphics[width=\linewidth]{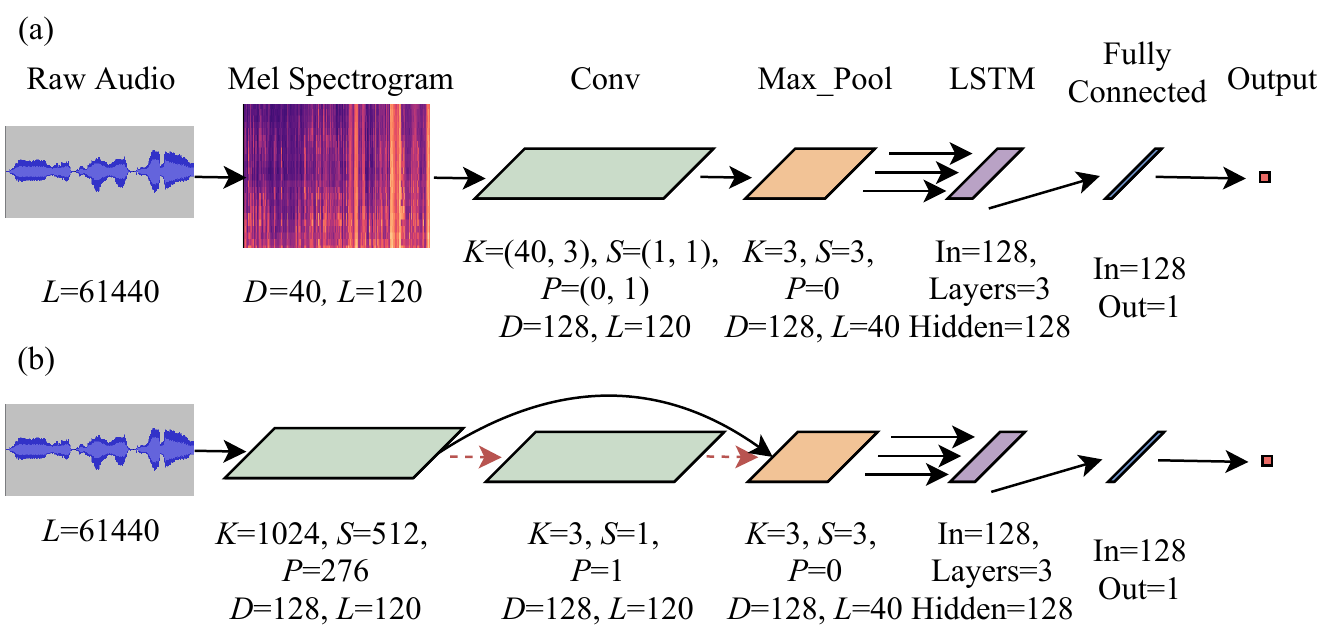}
  \caption{Network architectures, $D$ is depth (number of feature channels), $L$ is length, $K$ is the kernel size, $S$ is the stride length, and $P$ is the zero padding. (a) shows the DepAudioNet architecture (after \cite{ma_depaudionet:_2016}); (b) shows our raw audio model where the dotted red arrow shows an alternative network architecture with an additional convolutional filter.}
  \label{fig:both_models}
\end{figure}

Figure \ref{fig:both_models}(a) shows the overall DepAudioNet architecture. The source code for DepAudioNet is not available and so we reproduced the model through details given in the paper \cite{ma_depaudionet:_2016} using reasonable assumptions for unspecified aspects.

We experimented on the validation data to set the initial learning rate (LR) and rate of decay of the LR ($\lambda$) as no details in \cite{ma_depaudionet:_2016} were given. We initialised the LR to $0.001$ and decayed the LR by a factor of $0.9$ every $\lambda_{\rm{epoch}}$ equal to $2$ or $3$. We used the Adam optimiser, the F1-Score to measure the accuracy of the model, and the Binary Cross Entropy Loss to update the network weights.  


\subsection{DepAudioNet Reproduction Results}
\label{subsec:depaudionet reproduction}
As confirmation of successful reproduction, our reproduction achieved comparable results to those reported in \cite{ma_depaudionet:_2016}. As we can see from Table \ref{tab:baseline_depaudionet}, the performance of our reproductions with $\lambda=2$ and $\lambda=3$ are slightly better than the performance reported by Ma et al. \cite{ma_depaudionet:_2016}.

\section{Proposed Raw Audio Model}
\label{sec:proposed method}
We hypothesise that features such as the mel spectrogram used in DepAudioNet may emphasise pitch or vocal tract information that may be a proxy for gender, which could result in utilising the gender bias of DAIC-WOZ. To test the robustness of the mel-spectrogram features against the gender bias of DAIC-WOZ, we compared our reproduced DepAudioNet with an alternative raw audio model, Figure \ref{fig:both_models}(b). 


\begin{table}[t]
  \caption{Reported F1-Score results of several methods using the unaltered validation set of the DAIC-WOZ dataset. DepAnudioNet$^*$ is our reproduced DepAudioNet model based on Ma et al. \cite{ma_depaudionet:_2016}, $\lambda$ is the rate of decay of the LR in epochs and C is the number of convolutional filters.}
  \centering
  \label{tab:baseline_depaudionet}
  \begin{tabular}{| c | c | c | c | c | c |} 
  \hline
   \textbf{Model} & \boldsymbol{$\lambda$} & \textbf{C} & \textbf{F1 (ND)} & \textbf{F1 (D)} & \textbf{F1 avg} \\ 
   \hline
   DepAudioNet \cite{ma_depaudionet:_2016} & $-$ & $1$ & $.700$ & $.520$ & $.610$\\
   \hline
   DepAudioNet$^*$ & $2$ & $1$  & $.732$ & $.522$ & $.627$\\
   \hline
   DepAudioNet$^*$ & $3$ & $1$ & $.740$ & \textbf{.539} & $.634$ \\
   \hline
   Raw Audio & $2$ & $1$  & $.707$ & $.501$ & $.603$ \\ 
   \hline
   Raw Audio & $3$ & $1$ & $.707$ & $.531$ & $.619$ \\
   \hline
   Raw Audio & $2$ & $2$ & $.766$ & $.531$ & $.648$ \\ 
   \hline
   \textbf{Raw Audio} & \textbf{3} & \textbf{2} & \textbf{.796} & $.520$ & \textbf{.658} \\
  \hline
  \end{tabular}
\end{table}

Our proposed raw audio model follows the pre-processing outlined in Section \ref{subsec:File Processing} except that we use raw audio as our input instead of extracting a mel-spectrogram. The input size of the raw audio features is calculated according to \(r=N_{\rm{seg}}\times h\).

The kernel of the convolutional filter for our raw audio model matched the window and hop size from the mel spectrogram analysis from Section \ref{subsec:File Processing}, \(\{K=1024,\ S=512,\ P=276\}\), where $K$ is the kernel size, $S$ is the stride length, and $P$ is the zero padding. These values were chosen so that the output dimensions of the convolutional filter would match those used by Ma et al. \cite{ma_depaudionet:_2016}. We also experimented with an additional convolutional filter (shown by red dotted arrow in Figure \ref{fig:both_models}(b)) added after the first with parameters \(\{K=3,\ S=1,\ P=1\}\) (a 1D version of the convolutional filter in DepAudioNet shown in Figure \ref{fig:both_models}(a)). 

\subsection{Results of Proposed Raw Audio Model}
\label{subsec:results of proposed method}

From Table \ref{tab:baseline_depaudionet}, our proposed raw audio model with a single convolutional filter performed comparably, within $\pm{1\%}$, to the results reported by Ma et al. \cite{ma_depaudionet:_2016}. When we added an extra convolutional layer, our proposed raw audio model surpassed the performance of Ma et al. \cite{ma_depaudionet:_2016} and our reproduced DepAudioNet models.

In the next section we will evaluate the effect of the gender bias of the DAIC-WOZ and show how using raw audio as input adds robustness against this type of bias.

\begin{table*}[ht]
  \caption{Results on the validation set of our reproduced and proposed models showing the effect of gender balancing by utilising the sub-sampling data re-distribution method from \cite{krawczyk_learning_2016}. DepAudioNet$^*$ is our reproduced DepAudioNet model based on Ma et al. \cite{ma_depaudionet:_2016}, $\lambda$ is the rate of decay of the LR in epochs and C is the number of convolutional filters in the model.}
  \centering
  \label{tab:gender split results}
  \begin{tabular}{| c | c | c | c | c | c | c | c | c | c | c | c | c |}
  \hline
    \multirow{2}{*}{\textbf{Row}} & \multirow{2}{*}{\textbf{Model}} & \multirow{2}{*}{\boldsymbol{$\lambda$}} & \multirow{2}{*}{\textbf{C}} & \textbf{Gender} & \multicolumn{3}{|c|}{\textbf{Female}} & \multicolumn{3}{|c|}{\textbf{Male}} & \textbf{F1 Total} & \textbf{Difference} \\
    \cline{6-11} & & & & \textbf{Balance} & \textbf{\textit{F1 avg}} & \textbf{\textit{F1 (ND)}} & \textbf{\textit{F1 (D)}} & \textbf{\textit{F1 avg}} & \textbf{\textit{F1 (ND)}} & \textbf{\textit{F1 (D)}} & \textbf{Average} & \textbf{(\%)} \\
   \hline
   1 & DepAudioNet$^*$ & $2$ & $1$ & N & $.628$ & $.637$ & $.619$ & $.539$ & $.800$ & $.279$ & $.627$ & \multirow{2}{*}{$-14.0$}\\ 
   \cline{1-12}
   2 & DepAudioNet$^*$ & $2$ & $1$ & Y & $.547$ & $.696$ & $.397$ & $.497$ & $.730$ & $.264$ & $.539$ & \\
   \hline
   3 & DepAudioNet$^*$ & $3$ & $1$ & N & $.665$ & $.688$ & $.641$ & $.527$ & $.777$ & $.277$ & $.634$ & \multirow{2}{*}{$-13.4$} \\
   \cline{1-12}
   4 & DepAudioNet$^*$ & $3$ & $1$ & Y & $.573$ & $.697$ & $.449$ & $.514$ & $.703$ & $.324$ & $.549$ & \\
   \hline
   \hline
   5 & Raw Audio & $2$ & $1$ & N & $.546$ & $.555$ & $.537$ & $.618$ & $.819$ & $.417$ & $.604$ & \multirow{2}{*}{$+0.82$}\\ 
   \cline{1-12}
   6 & Raw Audio & $2$ & $1$ & Y & $.608$ & $.747$ & $.470$ & $.606$ & $.710$ & $.502$ & $.609$ & \\
   \hline
   7 & Raw Audio & $3$ & $1$ & N & $.526$ & $.505$ & $.547$ & $.659$ & $.846$ & $.471$ & $.619$ & \multirow{2}{*}{$+3.43$} \\
   \cline{1-12}
   8 & Raw Audio & $3$ & $1$ & Y & $.613$ & $.808$ & $.418$ & $.660$ & $.766$ & $.555$ & $.641$ & \\
   \hline
   \hline
   9 & Raw Audio & $2$ & $2$ & N & $.627$ & $.672$ & $.582$ & $.614$ & $.841$ & $.387$ & $.648$ & \multirow{2}{*}{$-4.78$} \\ 
   \cline{1-12}
   10 & Raw Audio & $2$ & $2$ & Y & $.600$ & $.782$ & $.419$ & $.635$ & $.763$ & $.507$ & $.617$ & \\
   \hline
   11 & Raw Audio & $3$ & $2$ & N & $.679$ & $.743$ & $.615$ & $.553$ & $.840$ & $.267$ & $.658$ & \multirow{2}{*}{$-4.26$} \\
   \cline{1-12}
   12 & Raw Audio & $3$ & $2$ & Y & $.586$ & $.711$ & $.462$ & $.668$ & $.737$ & $.598$ & $.630$ & \\
   \hline
  \end{tabular}
\end{table*}

\section{Impact of Gender Bias}
\label{sec:experimentation with gender balance}

We saw in Table \ref{tab:training_quadrants_train} (Section \ref{subsec:daic-woz dataset}) that the number of training examples varies according to class (D:ND) and gender (female/male). While Ma et al. \cite{ma_depaudionet:_2016} addressed the (D:ND) class imbalance by sub-sampling the ND examples, they did not report on gender bias. 

Following the data re-distribution approach of Krawczyk \cite{krawczyk_learning_2016} we performed data sub-sampling to have equal data in all quadrants of Table \ref{tab:training_quadrants_train}. We split the training data into the four quadrants in Table \ref{tab:training_quadrants_train}: (f, D), (f, ND), (m, D), (m, ND). We then randomly sub-sampled the audio signals from every quadrant with respect to the quadrant with the fewest examples, in this case $14$ audio signals from depressed males (m, D). Following this process, our training data equally represents every quadrant found in Table \ref{tab:training_quadrants_train} and ensures independence \cite{barocas-hardt-narayanan} from gender, such that \(p(D\ |\ g=\rm{f}\)\() = p(D\ |\ g=\rm{m}\)\()\) from the perspective of the training data. 

\subsection{Results of Gender Balance}
\label{subsec:results of gender balance}

The results after training the models following gender balance are shown in Table \ref{tab:gender split results}. The performance of DepAudioNet$^*$ (our reproduction of Ma et al. \cite{ma_depaudionet:_2016}) (rows 1 and 3 respectively) drops by $14.0\%$ and $13.4\%$ respectively after we employ gender balance (rows 2 and 4 respectively) from $.627$ to $.539$ (row 1 to row 2), and $.634$ to $.549$ in (row 3 to row 4). 

Our results using the proposed raw audio model vary. While using one convolutional filter (rows 5 and 7 respectively), we see a performance boost of $0.82\%$ and $3.43\%$ respectively (rows 6 and 8 respectively) after applying gender balance. But the addition of the second convolutional filter (rows 9 and 11 respectively) shows a drop in performance of $4.78\%$ and $4.36\%$ respectively after gender balance (rows 10 and  12 respectively), albeit less than the drop found using DepAudioNet$^*$. 

We can see (rows 1 to 2 and rows 3 to 4) that the (f, D) performance of DepAudioNet$^*$ drops more than our raw audio model (rows 5 to 6, 7 to 8, 9 to 10, and 11 to 12). 

The performance drop of DepAudioNet$^*$ suggests that DepAudioNet$^*$ may use gender-based information. Finally, we can see that performance on the minority class (m, D) increases across all models (with the exception of rows 1 to 2). Our proposed raw audio model shows superior results to DepAudioNet$^*$ in terms of overall performance and performance difference after applying gender balancing. These results suggest that raw audio may be more robust to gender bias than using the mel-spectrogram for this task. 

We recognise the limitations of this study as we have only tested the robustness of the mel-spectrogram and raw audio on one dataset and one model architecture. However, as a proof of concept our results suggest that further analysis would be beneficial to concretely determine the effect of gender bias on various datasets and input features. 

\section{Conclusion}
\label{sec:conclusion}
We have studied potential gender bias in the DAIC-WOZ dataset. In Section \ref{sec:experimentation with gender balance} we removed gender bias through data re-distribution by balancing the training data quadrants from Table \ref{tab:training_quadrants_train}. Data re-distribution helps the classifier to learn in an unbiased manner by ensuring independence from gender with respect to the training data. 

We showed in Table \ref{tab:gender split results} that leaving the dataset unaltered can result in overreporting performance due to the gender bias (shown by performance drops after applying gender balancing techniques), especially when using the mel-spectrogram. 

We found that using raw audio can provide a classification that is more robust to gender bias than the mel-spectrogram as shown by the performance changes before and after gender balancing. Using raw audio can also outperform results using the mel-spectrogram despite the network architecture being optimised for the mel-spectrogram and not raw audio. 

In future work, it would be interesting to explore other datasets which exhibit gender bias, and the robustness of different features against dataset bias. In addition, it would be useful to examine the robustness of other popular hand-crafted features such as the Mel-Frequency Cepstral Coefficients (MFCC). We would like to determine the efficacy of approaches such as gender normalisation on the gendered subsets of DAIC-WOZ \cite{huang_natural_2020} to tackle the gender bias of DAIC-WOZ as well.

Our code is available for download\footnote{https://github.com/adbailey1/DepAudioNet\_reproduction}.

\section*{Acknowledgement}
This work was funded by the Engineering and Physical Sciences Research Council (EPSRC) Doctoral Training Grant ``EP/R513350''.

\bibliographystyle{IEEEbib}
\bibliography{refs}

\end{document}